\documentclass[options]{JHEP3}

\title{Tunneling of massive and charged particles from noncommutative Reissner-Nordstr\"{o}m black hole}

\author{Kourosh Nozari\footnote{knozari@umz.ac.ir} \quad and \quad Sara Islamzadeh\footnote{s.islamzadeh@stu.umz.ac.ir} \\

Department of Physics, Faculty of Basic Sciences,
University of Mazandaran,\\
P. O. Box 47416-95447, Babolsar, IRAN}

\abstract{Massive charged and uncharged particles tunneling from
commutative Reissner-Nordstr\"{o}m black hole horizon has been
studied with details in literature. Here, by adopting the coherent
state picture of spacetime noncommutativity, we study tunneling of
massive and charged particles from a \emph{noncommutative} inspired
Reissner-Nordstr\"{o}m black hole horizon. We show that Hawking
radiation in this case is not purely thermal and there are
correlations between emitted modes. These correlations may provide a
solution to the information loss problem. We also study
thermodynamics of noncommutative horizon in this setup.}

\keywords{Black Hole, Noncommutative Geometry, Quantum Tunneling}

\begin{document}

\section{Introduction}

After the discovery of Hawking radiation \cite{1}, a lot of attempts
have been made to explore different aspects of this revolutionary
achievement. There are some important questions in this regard: Is
black hole radiation purely thermal? Are unitary and Lorentz
invariance symmetries preserved at the quantum gravity level? What
happens in the final stages of black hole evaporation? Are the
information that were entered horizon at the time of star formation
missing? Although, there is no perfect theory by now to answer these
questions properly, various methods are presented to address such
questions in recent years. One of these attempts is the strategy
provided by Parikh and Wilczek \cite{2}. In this approach, particle
and antiparticle pairs are created and the particle tunnels through
event horizon. Due to emission of this particle, total energy of
black hole reduces. Conservation of energy requires that the event
horizon radius reduces too. To deduce the black hole event horizon
thermodynamics, Parikh and Wilczek utilized the WKB approximation.
This approximation is actually justified since there is an infinite
blue shift in the vicinity of the horizon. Indeed, the barrier
through which tunneling occurs is induced by the emitted particle
itself. They have considered the tunneling particle as a spherically
symmetric shell that is ejected from black hole surface. This
approach was the basis of a lot of research programs then after.
Tunneling of massless \cite{3} and massive \cite{4} particles from
Schwarzschild black hole and also noncommutative inspired
Schwarzschild black hole \cite{5} are studied. Recently, tunneling
of massive and charged particles from Reissner-Nordstr\"{o}m black
hole horizon is reported too \cite{6}. Also extensions to higher
dimensional spacetime models are considered by some authors
\cite{7}.

In recent years noncommutative quantum field theory has been
attracted much attentions \cite{8}. Noncommutativity is an intrinsic
characteristic of manifold that implies the existence of a natural
ultra-violet cutoff (or equivalently a minimal measurable length) in
quantum field theory. Spacetime noncommutativity can be achieved
naturally on certain backgrounds of string theory. In this
viewpoint, description of the spacetime as a smooth commutative
manifold becomes therefore a mathematical assumption no more
justified by physics. It is then natural to relax this assumption
and conceive a more general noncommutative spacetime, where
uncertainty relations and spacetime discretion naturally arise. The
noncommutativity of spacetime can be encoded in the commutator
\begin{equation}
[\hat{x}^i,\hat{x}^j]=i\theta^{ij}
\end{equation}
where $\theta^{ij}$ is a real, antisymmetric and constant tensor,
which determines the fundamental cell discretion of spacetime much
in the same way as the Planck constant $\hbar$ discretizes the phase
space. This was motivated by the need to control the divergences
showing up in theories such as quantum electrodynamics. It has been
shown that noncommutativity eliminates point-like structures in the
favor of smeared objects in flat spacetime. As the authors have
shown in Ref. \cite{9}, the effect of smearing is mathematically
implemented as a substitution rule: position Dirac-delta function is
replaced everywhere with a Gaussian distribution of minimal width
$\sqrt{\theta}$. In this framework, they have chosen the mass
density of a static, spherically symmetric, smeared, particle-like
gravitational source as follows
\begin{equation}
\rho_\theta(r)=\frac{M}{(2\pi\theta)^{\frac{3}{2}}}\exp(-\frac{r^2}{4\theta}).
\end{equation}
As they have indicated, the particle mass $M$ (or particle charge
$Q$), instead of being perfectly localized at a point, is diffused
throughout a region of linear size $\sqrt{\theta}$. This is due to
the intrinsic uncertainty as has been shown in the coordinate
commutators (1.1). This view point on spacetime noncommutativity has
stimulated a lot of research programs in recent years, some of which
can be seen in Refs \cite{5,7,10}. In this paper we adopt this
viewpoint to study tunneling of massive and charged particles from a
noncommutative inspired Reissner-Nordstr\"{o}m black hole horizon.
We show that Hawking radiation in this case is not purely thermal
and there are correlations between emitted modes that may provide a
trace to address the information loss paradox. We study also
thermodynamics of noncommutative horizon in this setup. Finally we
investigate possible relation between charge and spacetime
noncommutativity in this setup.


\section{Motion of tunneling particles}
In this section we consider Einstein field equations in coherent
state noncommutative picture and we apply Peinlev\'{e}
transformation to derive equations of motion of massive and charged
particle in this setup. In coherent state picture of spacetime
noncommutativity, a particle of mass $M$ and charge $Q$, instead of
being perfectly localized at a point, is diffused throughout a
region of linear size $\sqrt{\theta}$. In fact, the Dirac-delta
function for pointlike structures is replaced everywhere with a
Gaussian distribution of minimal width $\sqrt{\theta}$. In this
framework, the mass and charge densities of a static, spherically
symmetric, smeared, particle-like gravitational source can be
modeled as follows

$$\rho_{mat}(r)=\frac{M}{(4\pi\theta)^{3/2}}e^{-(\frac{r^{2}}{4\theta})}$$
$$\rho_{el}(r)=\frac{Q}{(4\pi\theta)^{3/2}}e^{-(\frac{r^{2}}{4\theta})}$$
In coordinate coherent state approach, noncommutative effects enters
to energy-momentum tensor and current densities. Therefore, the
Einstein-Maxwell field equations are preserved in usual form
\begin{equation}
R^{\mu}_{\nu}-\frac{1}{2}\delta^{\mu}_{\
\nu}{R}=8\pi(T^{\mu}_{\nu}|_{mat}+T^{\mu}_{\nu}|_{el})
\end{equation}
\begin{equation}
\frac{1}{\sqrt{-g}}\partial_{\mu}\big(\sqrt{-g}F{^{\mu\nu}}\big)=J^{\nu}
\end{equation}
Solving Einstein equations (2.1), the line element for a spherically
symmetric system and the vector potential are as follows
\begin{equation}
ds^{2}=-g_{00}dt^{2}+g_{00}^{-1}dr^{2}+r^{2}d\Omega^{2}
\end{equation}
\begin{equation}
A_{\mu}=h(r)\delta_{\mu}^{0}
\end{equation}
with
$$g_{00}=1-\frac{2M_{\theta}}{r}+\frac{Q_{\theta}^{2}}{r^{2}}$$ and $$h(r)=-\frac{Q_{\theta}}{r}$$
where by definition
$$M_{\theta}(r)=\frac{2M}{\sqrt{\pi}}\gamma(\frac{3}{2},\frac{r^{2}}{4\theta})$$
$$Q_{\theta}(r)=\frac{Q}{\sqrt{\pi}}\sqrt{\gamma^{2}(\frac{1}{2},\frac{r^{2}}{4\theta})-
\frac{r}{\sqrt{2\theta}}\gamma(\frac{1}{2},\frac{r^{2}}{2\theta})}\,.$$
$\gamma$ in the above equations is incomplete lower Gamma function.
The line element (2.3) has singularity at the horizons. To describe
the tunneling process, we require a metric that is not singular on
the horizon. The Painlev\'{e} coordinates transformation is a
suitable tool to overcome this difficulty. With the Painlev\'{e}
transformation \cite{11} the metric (2.3) and vector potential (2.4)
take the following forms respectively
\begin{equation}
ds^{2}=-g_{00}dt_{p}^{2}+dr^{2}+2\sqrt{1-g_{00}}dt_{p}dr+r^{2}d\Omega^{2}
\end{equation}
\begin{equation}
A_{\mu}=h(r)\delta_{\mu}^{0}-h(r)\frac{\sqrt{1-g_{00}}}{g_{00}}\delta_{\mu}^{1}
\end{equation}
With this transformation, horizon's singularity is removed and we
can analyze tunneling process of particles through horizon. The
Lagrangian equation of motion of a particle with mass $m$ and charge
$q$ is obtained as follows
\begin{equation}
\mathcal{L}=\frac{m}{2}(-g_{00}\dot{t}_{p}^{2}+\dot{r}^{2}+2\sqrt{1-g_{00}}\dot{t}_{p}\dot{r})+qh(r)\dot{t}_{p}-
\frac{qh(r)\sqrt{1-g_{00}}}{g_{00}}\dot{r}
\end{equation}
where a dot indicates the derivative with respect
to proper time $\tau$.\\
By using the Euler-Lagrange equation ($\frac{\partial
\mathcal{L}}{\partial q}-\frac{d}{dt}\frac{\partial
\mathcal{L}}{\partial\dot{q}}=0$) we find
\begin{equation}
mg_{00}\dot{t}_{p}-m\sqrt{1-g_{00}}\dot{r}-qh(r)\equiv{\cal{E}}=constant
\end{equation}
To achieve the radial equation of motion of in the Painlev\'{e}
coordinates, we need timelike trajectories that are given by
\begin{equation}
g_{00}\dot{t}_{p}^{2}+\dot{r}^{2}+2\sqrt{1-g_{00}}\dot{t}_{p}\dot{r}=-1
\end{equation}
Equations (2.8) and (2.9) can be solved simultaneously to obtain
$\dot{r}$ and $\dot{t}$. Then equation of motion of a massive and
charged particle in Peinlev\'{e} coordinates is obtained as follows
\begin{equation}
\frac{dr}{dt_{p}}=\pm
g_{00}\frac{\sqrt{(qh(r)+{\cal{E}})^{2}-m^{2}g_{00}}}{qh(r)+{\cal{E}}\pm\sqrt{1-g_{00}}\sqrt{(qh(r)+{\cal{E}})^{2}-m^{2}g_{00}}}
\end{equation}
This equation is the basis of our forthcoming analysis.

\section{Thermodynamics of noncommutative Reissner-Nordestr\"{o}m black hole}

In this section we study tunneling of massive and charged particles
through horizon of a Reissner-Nordstr\"{o}m black hole in a
noncommutative space. Reissner-Nordstr\"{o}m black hole can radiate
through two ways

1- By radiating neutral particles. It causes the black hole to reach
a Schwarzschild black hole in the final stage of evaporation.

2- By radiating charged particles. This happens through creation of
charged particle and antipartile pairs.

We note that the second process leads to a black hole extreme
configuration that is called the Reissner Nordstr\"{o}m black hole
remnant. As has been indicated in Ref. \cite{12}, electric field at
the horizon in noncommutative picture is larger in value than the
critical electric field and so black hole is able to create
pair-charged particles. In which follows we focus on the later
picture, that is, we consider radiation of pair-charged particles.
As has been mentioned above, there are infinity blue shift in the
vicinity of black hole event horizon. So, we can use the WKB
approximation and calculate coefficients of transmission for a
massive and charged particle that tunnels from inner radius to the
outer radius. First, we consider imaginary part of the action for a
particle that is coming from an initial state with $r_{in}$ to a
final state with $r_{out}$
$$Im\:S\equiv Im\int
E\:dt=Im\int_{r_{in}}^{r_{out}}
p_{r}\:dr=Im\int_{r_{in}}^{r_{out}}\int_0^{p_{r}}\:dp_r\:dr$$ Using
the Hamilton equation of motion, $dp_r=\frac{dH}{\dot{r}}$, we find
\begin{equation}
Im\:S=Im\int_{r_{in}}^{r_{out}}\int_m^{\cal{E}}\frac{dH}{\dot{r}}\:dr=
-Im\int_m^{\cal{E}}\int_{r_{in}}^{r_{out}}\frac{dr}{\dot{r}}\:d\tilde{{\cal{E}}}
\end{equation}
Since the emitted particles are assumed to be massive, the lower
limit of integral now is $m$ instead of being zero as for massless
particles. Indeed, before tunneling, spherical shell of particle has
an energy ${\cal{E}}$ and after crossing the event horizon, changed
its energy to ${\cal{E}}-m$ and we'll represent effect of $m$ on the
tunneling rate. In previous section we probed motion of a massive
and charged particle in Painlev\'{e}'s coordinates and we achieved
equation of motion as given by Eq. (2.10). By expanding the metric
around the event horizon,\,
$g_{00}(r)=g_{00}(r_{\theta+})+g'_{00}(r_{\theta+})(r-r_{\theta+})+...,$\,
Eq. (2.10) takes the following form
\begin{equation}
\dot{r}=\pm g'(r_{\theta+})(r-r_{\theta+})\frac{\sqrt{(q
h(r)+{\cal{E}})^{2}-m^{2}g'(r_{\theta+})(r-r_{\theta+})}}{(q
h(r)+{\cal{E}})\pm\sqrt{1-g'(r_{\theta+})(r-r_{\theta+})}\sqrt{(q
h(r)+{\cal{E}})^{2}-m^{2}g'(r_{\theta+})(r-r_{\theta+})}}
\end{equation}
By substituting (3.2) in Eq. (3.1), integral has a pole at
$r=r_\theta$. We solve this integral by using the calculus of
residues. We find
$$Im\:S=\pi\int_m^{\cal{E}}
\frac{2}{g'_{00}(r_{\theta+})}\:d\tilde{{\cal{E}}}$$ By
considering\, $g_{00}=\frac{(r-r_{\theta+})(r-r_{\theta-})}{r^2}$,\,
the imaginary part of the action takes the following form
\begin{equation}
Im\:S=\pi\int_{m}^{{\cal{E}}}\frac{2
r_{\theta+}}{r_{\theta+}-r_{\theta-}}\:d\tilde{{\cal{E}}}
\end{equation}
To solve this integral we need to find the inner and outer
Reissner-Nordstrom's event horizons in noncommutative geometry. So
we $g_{00}(r_{\theta+})=0$ to find
$$r_{\theta\pm}(r_{\theta\pm})=M_{\theta}(r_{\theta\pm})\pm\sqrt{M_{\theta}^2(r_\theta\pm)-Q_{\theta}^2(r_\theta\pm)}\,.$$
This equation has no analytical solution for $r_{\theta\pm}$. We can
replace $r_{\theta\pm}$ by $r_{\pm}$ in lower incomplete Gamma
function \cite{13} to obtain the approximate event horizon radius as
\begin{equation}
r_{\theta\pm}(r_{{\theta\pm}})\cong
r_{\theta\pm}(M,Q)=\mathcal{M}_{\theta\pm}(M,Q)\pm\sqrt{\mathcal{M}_{\theta\pm}^2(M,Q)-\mathcal{Q}_{\theta\pm}^2(M,Q)}
\end{equation}
So we find \\
$$\mathcal{M}_{\theta\pm}(M,Q)=M\bigg[erf\Big(\frac{M\pm\sqrt{M^2-Q^2}}{2\sqrt{\theta}}\Big)-
\frac{M\pm\sqrt{M^2-Q^2}}{\sqrt{\pi\theta}}\exp\Big(-\frac{(M\pm\sqrt{M^2-Q^2})^2}{4\theta}\Big)\bigg]$$
$$\mathcal{Q}_{\theta\pm}(M,Q)=Q\sqrt{erf^2\Big(\frac{M\pm\sqrt{M^2-Q^2}}{2\sqrt{\theta}}\Big)
-\frac{M\pm\sqrt{M^2-Q^2}}{\sqrt{2\pi\theta}}erf\Big(\frac{M\pm\sqrt{M^2-Q^2}}{\sqrt{2\theta}}\Big)}\,$$
where $erf(x)$ is the \emph{Error Function}. So, $r_{\theta\pm}$ in
integral (3.3) becomes
\begin{equation}
r_{\theta\pm}=\mathcal{M}_{\theta\pm}(M-\tilde{{\cal{E}}},Q-q)\pm
\sqrt{\mathcal{M}_{\theta\pm}^2(M-\tilde{{\cal{E}}},Q-q)-\mathcal{Q}_{\theta+}^2(M-\tilde{{\cal{E}}},Q-q)}
\end{equation}
We note also that, $r_{in}$ and $r_{out}$ are obtained as follows
$$r_{in}=\mathcal{M}_{\theta+}(M,Q)+\sqrt{\mathcal{M}_{\theta+}^2(M,Q)-\mathcal{Q}_{\theta+}^2(M,Q)}$$
$$r_{out}=\mathcal{M}_{\theta+}(M-{\cal{E}},Q-q)+\sqrt{\mathcal{M}_{\theta+}^2(M-{\cal{E}},Q-q)-\mathcal{Q}_{\theta+}^2(M-{\cal{E}},Q-q)}$$
and the integral pole lies between these to extremes. Substituting
(3.5) into (3.6), the final form of the integral that should be
calculated is as follows
\begin{equation}
Im\:S=\int_m^{\cal{E}}
\frac{2\mathcal{M}_{\theta+}^{2}(M-\tilde{{\cal{E}}},Q-q)-
\mathcal{Q}_{\theta+}^{2}(M-\tilde{{\cal{E}}},Q-q)}{\sqrt{\mathcal{M}_{\theta+}^{2}(M-\tilde{{\cal{E}}},Q-q)-
\mathcal{Q}_{\theta+}^{2}(M-\tilde{{\cal{E}}},Q-q)}}\:d\tilde{{\cal{E}}}+2\int_m^{\cal{E}}
\mathcal{M}_{\theta+}^{2}(M-\tilde{{\cal{E}}},Q-q)\:d\tilde{{\cal{E}}}
\end{equation}
Analytical and exact solutions of the above integral is impossible.
Thus, while we try to preserve noncommutativity effects, we expand
the Gamma function around $\tilde{{\cal{E}}}$ to find some
analytical results. In this situation, $\mathcal{M}_{\theta}$ and
$\mathcal{Q}_{\theta}$ take the following forms respectively

\begin{equation}
\mathcal{M}_{\theta+}(M-\tilde{\cal{E}},Q-q)\simeq
(M-\tilde{\cal{E}})\eta
\end{equation}
and
\begin{equation}
\mathcal{Q}_{\theta+}(M-\tilde{\cal{E}},Q-q)\simeq (Q-q)\zeta
\end{equation}
where by definition,
\begin{equation}
\eta=\frac{2}{\sqrt{\pi}}\gamma\Big(\frac{3}{2},\frac{(M+\sqrt{M^2-Q^2})^2}{4\theta}\Big)
\end{equation}
and
\begin{equation}
\zeta=\frac{1}{\sqrt{\pi}}\sqrt{\gamma^2\Big(\frac{1}{2},\,
\frac{(M+\sqrt{M^2-Q^2})^2}{4\theta}\Big)-
\frac{M+\sqrt{M^2-Q^2}}{\sqrt{2\theta}}\:\gamma\Big(\frac{1}{2},\,
\frac{(M+\sqrt{M^2-Q^2})^2}{2\theta}\Big)}
\end{equation}
Eventually, substituting Eqs. (3.7) and (3.8) into Eq. (3.6) we
obtain the imaginary part of the action as follows
\begin{eqnarray}
Im\:S&=&\pi\:\Big[\eta\:(M-m)^2-\eta\:(M-{\cal{E}})^2\nonumber\\&+&(M-m)\:\sqrt{\eta^2\:(M-m)^2-\zeta^2\:(Q-q)^2}\nonumber
\\&-&(M-{\cal{E}})\:\sqrt{\eta^2\:(M-{\cal{E}})^2-\zeta^2\:(Q-q)^2}\Big]
\end{eqnarray}
Since $\Gamma=e^{-\frac{\cal{E}}{T}}\sim e^{-2Im\:S}$, existence of
nonlinear terms in the imaginary part of the action requires
correlation between the emitted modes. In other words, now two
different particles with energies ${\cal{E}}_1$ and ${\cal{E}}_2$
are correlated. This means that
$\Gamma_{{\cal{E}}_1+{\cal{E}}_2}\neq\Gamma_{{\cal{E}}_1}+\Gamma_{{\cal{E}}_2}$.
Therefore, emission rate is no longer purely thermal and part of
information can be supplied in correlations between emitted modes.
Moreover, transition coefficient depends on energy, mass and charge
of emitted particle and also spacetime noncommutativity parameter.\\
To obtain temperature, we expand Eq. (3.11) with respect to
$m,\cal{E}$ and $q$ to find
\begin{eqnarray}
Im\:S&=&\pi\:\frac{\Big(\eta M +\sqrt{\eta^2 M^2-\zeta^2
Q^2}\Big)^2}{\sqrt{\eta^2 M^2-\zeta^2
Q^2}}\:\bigg\{\Big(1-\frac{q\zeta^{2}Q(3\eta^{2}M^{2}-\zeta^{2}Q^{2})}{(\eta^2
M^2-\zeta^2 Q^2)\:(\eta M +\sqrt{\eta^2 M^2-\zeta^2
Q^2})^2}\Big)\:({\cal{E}}-m)\nonumber\\&+&\frac{2\eta
(\eta^2M^2-\zeta^2Q^2)^\frac{3}{2}+2 \eta^4 M^3-3\eta^2 \zeta^2 M
Q^2}{(\eta^2 M^2-\zeta^2 Q^2)\:(\eta M+\sqrt{\eta^2 M^2-\zeta^2
Q^2})}\:(m^2-{\cal{E}}^{2})+...\bigg\}
\end{eqnarray}
We see that when the emitted particle is massive, the mass of the
particle appears explicitly in tunneling rate. Since spacetime at
infinity tends to the Minkowski spacetime, the observer at infinity
detects particles on-shell. Also, since ${\cal{E}}>m$, the condition
$Im S>0$ is satisfied. Therefore, the emitted particle mass causes a
shift in tunneling rate through noncommutative horizon. Now the
modified Hawking temperature for Reissner-Nordstr\"{o}m black hole
in noncommutative space is as follows
\begin{equation}
T=\frac{\sqrt{\eta^2 M^2-\zeta^2 Q^2}}{2\pi\Big((\eta M+\sqrt{\eta^2
M^2-\zeta^2 Q^2})^2-\zeta^2 q Q \frac{3\eta^{2}M^{2}-\zeta^2
Q^2}{\eta^2 M^2-\zeta^2 Q^2}\Big)}
\end{equation}
The effect of noncommutativity is hidden in $\eta$ and $\zeta$ that
are defined by Eqs. (3.9) and (3.10) respectively. Note that we
considered the general case of tunneling of massive and charged
particles though event horizon of a noncommutative
Reissner-Nordstr\"{o}m black hole. If we set $\theta=0$, temperature
reduces to the commutative case result that has been reported in
Ref. \cite{6} and if we set $q=0$, we receive the classical
commutative space Hawking temperature
$T_{H}=\frac{\sqrt{M^2-Q^2}}{2\pi (M+\sqrt{M^2-Q^2})^2}$. Also, our
result is consistent with noncommutative case with $Q=0$ that is
reported in Ref. \cite{5}. Indeed, our result contains all limiting
cases properly.\\

Figure 1 shows the Hawking temperature of a Reissner-Nordstr\"{o}m
black hole versus its mass in noncommutative space and for different
black hole charge. As we see, the final state temperature decreases
as the black hole charge increases. We note that black hole
evaporates through radiation of charged particle-antiparticle pairs
until it reaches a remnant with maximal temperature. In Ref.
\cite{12} analytical results are obtained for minimum mass of
Reissner-Nordstr\"{o}m black hole in noncommutative geometry. This
minimal mass is due to considering of Gaussian distribution in
noncommutative geometry for mass and charge. As figure 1
illustrates, temperature of Reissner-Nordstr\"{o}m black hole
remnant is less than the temperature of Schwarzschild black hole
remnant. So, the Reissner-Nordstr\"{o}m black hole remnant is colder
than the Schwarzschild black hole remnant. As another important
outcome, the noncommutativity effect becomes more effective for
small charges. We note that although our results agree with results
obtained in Ref. \cite{12}, but we obtained these results in a
different manner through tunneling method.\\

\begin{figure}[htp]
\begin{center}
\includegraphics{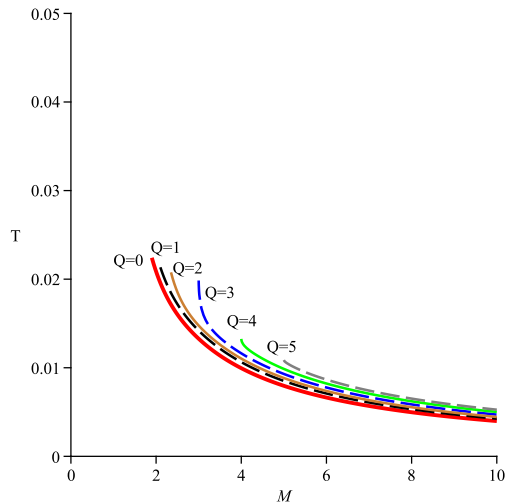}\vspace{7cm}
\caption{\scriptsize{Temperature of a \emph{noncommutative}
Reissner-Nordstr\"{o}m black hole versus its mass for different
charge.}}
\end{center}
\end{figure}
To compare with commutative case, we present figure 2 that is
obtained for a commutative Reissner-Nordstr\"{o}m black hole. The
authors in Ref. \cite{14} have shown that a \emph{noncommutative}
Schwarzschild black hole has features very similar to a
\emph{commutative} Reissner-Nordstr\"{o}m black hole. They have
concluded that there is a close relation between charge and
noncommutativity. Here we see that Reissner-Nordstr\"{o}m black hole
in commutative space has a remnant in the final stage of its
evaporation just due to its charge. In other words, charge by itself
prevents total evaporation of a Reissner-Nordstr\"{o}m black hole
and this is much similar to say that noncommutativity prevents a
Schwarzschild black hole from total evaporation. This shows that
existence of black hole \emph{charge} can address at least part of
the temperature divergence at final stage of evaporation and
information loss problems by prediction of a final state non-zero
mass remnant. We note also that the remnant mass for commutative and
noncommutative Reissner-Nordstr\"{o}m black hole are different. This
is the case also for temperature.

\begin{figure}[htp]
\begin{center}
\includegraphics{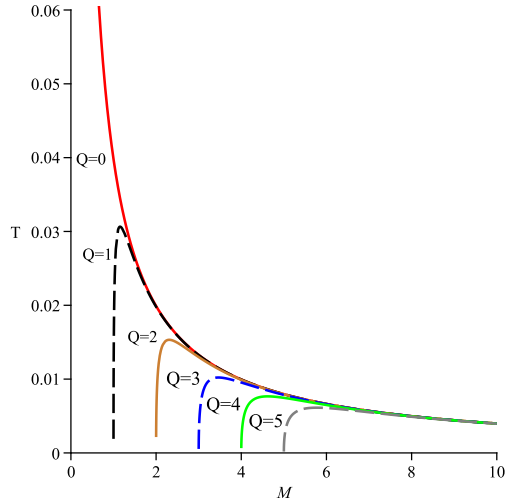}\vspace{7cm} \caption{\scriptsize{
Temperature of a \emph{commutative} Reissner-Nordstr\"{o}m black
hole versus its mass for different charges.}}
\end{center}
\end{figure}

\section{Conclusion}
Within the coherent state picture of spacetime noncommutativity, we
have studied tunneling of massive, charged particles in
noncommutative horizon of a Reissner-Nordstr\"{o}m black hole. We
applied the standard Parikh-Wilczek tunneling method by adopting the
Peinlev\'{e} transformation in order to remove singularities of
metric. We have shown that tunneling rate is dependent on the
energy, mass and charge of tunneling particles and also the
noncommutativity parameter $\theta$. We have shown that
noncommutative Reissner-Nordstr\"{o}m black hole radiates through
Hawking process until it reaches a remnant whose final temperature
and mass depends on the noncommutativity parameter and total charge
of the black hole. Also we have shown that the
Reissner-Nordstr\"{o}m black hole remnant is colder than the
Schwarzschild black hole remnant and the noncommutativity effect
becomes more effective for small black hole charges. Finally by
comparing evaporation process of a commutative
Reissner-Nordstr\"{o}m black hole with a noncommutative
Schwarzschild black hole we concluded that black hole charge by
itself can prevent total evaporation of Reissner-Nordstr\"{o}m black
hole much in the same way that spacetime noncommutativity prevents
total evaporation of a Schwarzschild black hole. This feature is in
agreement with previous finding in Ref. \cite{14} that there is some
nontrivial connection between charge and noncommutativity.

\end{document}